\documentclass[aps,preprintnumbers, superscriptaddress, amsmath, showpacs]{revtex4}
\usepackage{epsfig}
\usepackage{psfrag}
\usepackage{graphicx}% Include figure files
\usepackage{dcolumn}% Align table columns on decimal point
\usepackage{bm}% bold math

\begin{document}

\preprint{RU06-9-B}

\title{Angular Momentum Mixing in a Non-spherical Color Superconductor}% Force line breaks with \\

\author{Bo  Feng}
 \email{fengbo@iopp. ccnu. edu. cn}%Lines break automatically or can be forced with \\
\affiliation{Institute of Particle Physics,  Huazhong Normal University,
Wuhan,  430079,  China}
\author{ De-fu  Hou}%
 \email{hdf@iopp. ccnu. edu. cn}
 \affiliation{Institute of Particle Physics,  Huazhong Normal University,
Wuhan,  430079,  China}
%\author{Jia-rong Li}
%\email{ljr@iopp. ccnu. edu. cn} \affiliation{Institute of Particle
%Physics,  Huazhong Normal University, Wuhan,  430079,  China}
\author{Hai-cang Ren}
\email{ren@mail.rockefeller.edu} \affiliation{Physics Department, The
Rockefeller University, 1230 York Avenue,  New York,  NY
10021-6399} \affiliation{Institute of Particle Physics, Huazhong
Normal University, Wuhan,  430079,  China}

\date{\today}% It is always \today,  today,

\begin{abstract}
We study the angular momentum mixing effects in the color
superconductor with non-spherical pairing.  We first clarify the
concept of the angular momentum mixing with  a toy model for
non-relativistic and spinless fermions. Then we derive the gap 
equation for the polar phase of dense QCD by minimizing the 
CJT free energy. The solution of the gap equation consists of all 
angular momentum partial waves of odd parity. The corresponding 
free energy is found to be lower than that reported in the 
literature with $p$-wave only.
\end{abstract}

\pacs{26.30.+k, 91.65.Dt, 98.80.Ft}% PACS,  the Physics and Astronomy
\maketitle

\section{Introduction}\label{sec:level1}

The properties of quark matter at extreme conditions have been
an active research area  both theoretically and experimentally. At
high temperature, the quark-gluon plasma(QGP) has long been
searched by colliding two nuclei at sufficiently high energy. On
the other side, we expect that quark matter becomes color
superconducting through a phase transition at high baryon density
but low temperature\cite{B,DA,ARW1,RTEM,MKF,dk}, which is the
typical condition inside compact stars.

In a typical metallic superconductor, the electrons pair with
equal chemical potential near the Fermi surface. The situation
with a quark matter, however, is much more complicated. While the
quark-quark interaction favors pairing between quarks of different
flavors, the mass difference among $u$, $d$ and $s$ together with
the charge neutrality requirement induces a substantial mismatch
among their Fermi momenta at the baryon density inside a compact
star and thereby reduces the available phase space for Cooper
pairing. A number of exotic color superconductivity phases
in the presence of mismatch have been proposed in the
literature\cite{HZC,MI,MCK,ABR,BCR}, but a consensus point of view
of the true ground state has not been reached. The single flavor
pairing\cite{T,A,SSW,SWR,Rev}, which is free from the Fermi momentum
mismatch, is an interesting alternative in this circumstance and
will be considered here. Since the quark-quark interaction is
attractive in the color anti-triplet channel, the color wave
function of the pair is anti-symmetric. For the equal helicity pairing 
to be considered in this article, the parity of the orbital wave function has
to be odd as required by the Pauli principle. Except for the
color-spin-lock phase examined in \cite{T}, the energy gap will
not be spherical. The odd parity prevents the diquark wave function
from realizing the full pairing potential. The energy scale of the
color superconductivity is therefore reduced.

At ultra-high baryon densities, asymptotic freedom of QCD ensures
the validity of the weak coupling expansion, which has been
carried out for CSC by a number of
authors\cite{D,TF,RD,DVIL,WJH2,WJH,WJH3,Qun,dk,BDJH,P}. The
dominant pairing interaction is mediated by one-gluon exchange and
can be decomposed into partial waves, as is
shown in Eq.(\ref{33}) below. Quantitative results of the
transition temperature in the equal helicity channel with
an arbitrary angular momentum has been obtained from the first
principle \cite{WJH2,WJH}, and read
\begin{equation}
T_{c}^{(J)}=512\pi^3(\frac{2}{N_f})^{5/2}\frac{\mu}{g^5}\exp
\Big[-\frac{3\pi^2}{\sqrt{2}g}+\gamma
-\frac{1}{8}(\pi^2+4)+3c_J\Big] \label{1}
\end{equation}
where $N_f$ is the number of flavor, $N$ is the number of colors,
$\mu$ is the chemical potential, $g$ is the running coupling
constant of QCD and $\gamma(=0.5772...)$ is the Euler constant.
The $J$-dependent constant
\begin{equation}
c_J = \left\{
\begin{array}{ll}0,\hspace{1.5cm}  { \rm for}\hspace{0.2cm}
J=0,\\
\hfill \\
 -2\sum\limits_{n=1}^J\frac{1}{n} \hspace{0.5cm}  {\rm
for}\hspace{0.2cm} J>0.
\end{array}
\right.
\label{subleading}
\end{equation}
The equal helicity pairing of odd parity picks up $T_c^{(1)}$ as
the transition temperature. We have
$T_c^{(1)}=e^{-6}T_c^{(0)}\simeq 2.479 \times 10^{-3}T_c^{(0)}$.
Natural analogy is drawn with the superfluidity of $\rm He^3$. But
important difference between the pairing potential in quark matter
and that in $\rm He^3$ has to be considered before ascertaining
the angular dependence of the energy gap. The forward singularity
of the one-gluon exchange renders the pairing strength {\it equal}
for {\it all} partial waves (the same leading order term inside
the bracket of (\ref{1})). The paring potential in $\rm He^3$,
however, is entirely in the channel of $J=1$.

The transition temperature was determined from the pairing
instability of the diquark scattering amplitude in the normal
phase. In a perturbative treatment, the Dyson-Schwinger equation
for the scattering amplitude is linear and the partial wave
decomposition in Eq.(\ref{1}) is legitimate. This is not the case
with the gap equation below $T_c$, which is nonlinear. A
non-spherical solution of the gap equation, in general, will be a
mixture of different partial waves unless the the pairing is
dominated in one angular momentum channel. The adjective
"non-spherical" refers to the magnitude of the energy gap of the
quasi-particle. So the $s$-wave and the CSL gaps are spherical and
are therefore free from the angular momentum mixing. The gap
equations in these cases are linear with respect to the angular
dependence. A nonspherical gap function, $\phi_M$, for the equal
helicity pairing at $T<T_c$ contains all spherical harmonics of
odd parity with the same azimuthal quantum number $M$. We have
\begin{equation}
\phi_M=\phi_{\rm 2SC}f_M(\hat p) \label{gap2}
\end{equation}
where $\phi_{\rm 2SC}$ is the gap function of 2SC in the absence of the
mismatch, $\hat p$ the direction of the relative momentum of the two quarks
in a Cooper pair and the angular dependent factor
\begin{equation}
f_M(\hat p)=\sum_{J=1,3,5,...}b_JY_{JM}(\hat p)
\end{equation}
with $J$ the {\it total} angular momentum of the Cooper pair.
Carrying the formulation of $\rm He^3$ over to QCD amounts to drop
all higher multipoles except that of $J=1$, which will not satisfy
the gap equation of QCD. It was argued in the literature that
$b_1=O(1)$ but $b_J=O(g)$ for $J\neq 1$, This, as will be shown
below, is not the case. Instead, we find that the function
$f_M(\hat p)$ satisfies a nontrivial integral equation and thus
$b_J=O(1)$ for all odd $J$'s. Therefore the angular momentum
mixing does occurs in the subleading order of the gap function.
The angular momentum mixing will modify all non-spherical "spin-1"
CSC examined in the literature, we shall focus our attention in
this paper to the equal helicity pairing with zero azimuthal quantum number, 
i.e. the analog of the polar phase of $\rm He^3$. The subscript $M$ of
$\phi_M$ and $f_M$ will be suppressed below. Even though this
phase is unstable, it is the simplest one to illustrate the mixing
mechanism.

The current work is organized as follows. In the next section, we
shall clarify the concept of the angular momentum mixing with a
toy model of non-relativistic and spinless fermions. In the Sect.
III the gap equation for the single flavor CSC will be derived by
minimizing the CJT free energy of QCD. This gap equation will be
reduced to an nonlinear integral equation for the function $f(\hat
p)$ in the subsequent section and the numerical solution will be
presented in the Sect, V. We conclude the paper in the Sect. VI.
Some technical details are deferred to the Appendices.
Our units are $\hbar=c=k_B=1$ and 4-vectors are denoted
by capital letters, $K \equiv K^\mu = (k_0,\vec{k})$ with $k_0$
the Matsubara energy, which becomes continuous at $T_c=0$.
Throughout the article, we shall follow the definition of the
leading order and the subleading order in \cite{dk}. Upon taking
the logarithm of the transition temperature or the magnitude of
the gap function, the order $O(\frac{1}{g})$ will be referred to
as the leading one and the $O(1)$ term to the subleading one.

\section{A toy model with angular momentum mixing}

To clarify the concept of the angular momentum mixing, we consider
a toy model of nonrelativistic and spinless fermions. In terms of
the creation and annihilation operators, the model Hamiltonian
reads
\begin{equation}
H=\sum_{\vec p}\epsilon_pa_{\vec p}^\dagger a_{\vec p}
-\frac{\lambda}{4\Omega}{\sum_{\vec p,\vec p^\prime}}^\prime
V(\hat p\cdot\hat p^\prime)a_{\vec p}^\dagger a_{-\vec p}^\dagger
a_{-\vec p^\prime}a_{\vec p^\prime} \label{toy}
\end{equation}
where $\epsilon_p=\frac{p^2}{2m}-\mu$ with $m$ the mass and $\mu$ the
chemical potential, $\lambda>0$ is a coupling constant, $\Omega$
is the normalization volume and the summation $\sum_{\vec p,\vec
p^\prime}^\prime$ extends to states with $|\epsilon_p|<\omega_D$ and
$|\epsilon_{p^\prime}|<\omega_D$ with $\omega_D$ a UV cutoff (Debye frequency
for electronic superconductors). The angular dependent form factor
$V(\hat p\cdot\hat p^\prime)$ can be expanded in series of the
Legendre polynomials,
\begin{equation}
V(\hat p\cdot\hat p^\prime)=\sum_{J=0}^\infty(2J+1)v_JP_J(\hat p\cdot\hat p^\prime).
\label{partial}
\end{equation}

Introducing the order parameter
\begin{equation}
\chi(\vec p)=<|a_{-\vec p}a_{\vec p}|>
\end{equation}
with $|>$ the ground state and expanding the interaction term of
(\ref{toy}) to the linear order of the fluctuation $a_{-\vec
p}a_{\vec p}-<|a_{-\vec p}a_{\vec p}|>$, we end up with the
mean-field Hamiltonian
\begin{equation}
H_{\rm MF}=\frac{1}{2}\sum_{\vec p,|\epsilon_p|<\omega_D}\chi^*(\vec
p)\phi(\vec p) +\sum_{\vec p}\epsilon_p a_{\vec p}^\dagger a_{\vec p}
-\frac{1}{2}\sum_{\vec p,|\epsilon_p|<\omega_D}\Big[\phi^*(\vec
p)a_{-\vec p}a_{\vec p} +\phi(\vec p)a_{\vec p}^\dagger a_{-\vec
p}^\dagger\Big],
\end{equation}
where we have introduced the gap function via
\begin{equation}
\phi(\vec p)=\frac{\lambda}{2\Omega}\sum_{\vec
p^\prime,|\epsilon_{p^\prime}|<\omega_D} V(\hat p\cdot\vec
p^\prime)\chi(\vec p^\prime). \label{gapdef}
\end{equation}
We have $\chi(-\vec p)=-\chi(\vec p)$ and $\phi(-\vec
p)=-\phi(\vec p)$ following from their definitions. Upon a Bogoliubov
transformation, we find that
\begin{equation}
\chi(\vec p)=\frac{\phi(\vec p)}{2\varepsilon_p} \label{bogoliubov}
\end{equation}
with $\varepsilon_p=\sqrt{\epsilon_p^2+|\phi(\vec p)|^2}$ and the ground state
energy
\begin{equation}
E_0=\sum_{\vec p,\epsilon_p<0}\epsilon_p+\Omega F
\end{equation}
with $F$ the condensation energy density given by
\begin{equation}
F=\frac{1}{2}\sum_{\vec p,|\epsilon_p|<\omega_D}\Big[\frac{\phi^*(\vec
p)\phi(\vec p)} {2\varepsilon_p}+|\epsilon_p|-\varepsilon_p\Big]. \label{cond}
\end{equation}
Substituting (\ref{bogoliubov}) into (\ref{gapdef}), we obtain the gap equation
\begin{equation}
\phi(\vec p)=\frac{\lambda}{4\Omega}\sum_{\vec
p^\prime,|\epsilon_{p^\prime}|<\omega_D} V(\hat p\cdot\hat
p^\prime)\frac{\phi(\vec p^\prime)}{\varepsilon_{p^\prime}}. \label{gapeq}
\end{equation}

In the weak coupling limit, $\omega_D<<\mu$ and
$\lambda D_F<<1$ with $D_F=\frac{m^{\frac{3}{2}}\mu^{\frac{1}{2}}}{\sqrt{2}\pi^2}$
the density of states on the Fermi surface, but the magnitude of
$V(\hat p\cdot\hat p^\prime)$ remains of order one. We have
\begin{equation}
\sum_{\vec p,|\epsilon_p|<\omega_D}=\Omega\int_{|\epsilon_p|<\omega_D}
\frac{d^3\vec p}{(2\pi)^3}\simeq \Omega \frac{D_F}{4\pi}\int
d^2\hat p \int_{-\omega_D}^{\omega_D} d\epsilon. \label{weak}
\end{equation}
Also, the support of the gap function extends
only to a narrow band around the Fermi surface. We may ignore the
dependence of $\phi$ on the magnitude $p=|\vec p|$ and switch the
argument of $\phi$ from $\vec p$ to $\hat p$. Following
(\ref{weak}), the integrations over $p$ in (\ref{cond}) and
(\ref{gapeq}) can be carried out readily and we end up with
\begin{equation}
F=-\frac{D_F}{16\pi}\int d^2\hat p|\phi(\hat p)|^2.
\end{equation}
and
\begin{equation}
\phi(\hat p)=\frac{\lambda D_F}{8\pi}\int d^2\hat p^\prime V(\hat
p\cdot\hat p^\prime) \phi(\hat
p^\prime)\ln\frac{2\omega_D}{|\phi(\hat p^\prime)|}.
\label{gapeqwk}
\end{equation}
The gap equation (\ref{gapeqwk}) is nonlinear because of the
logarithm.

In what follows, we consider two extremes of $V(\hat p\cdot\hat p^\prime)$, each of
which gives rise to an exact solution to the gap equation (\ref{gapeqwk}). We present
only the solution that is invariant under time reversal, i.e. the one with zero
azimuthal quantum number.

\noindent
{\it {Case I}}:
\begin{equation}
V(\hat p\cdot\hat p^\prime)=3P_1(\hat p\cdot\hat p^\prime)=3\hat p\cdot\hat p^\prime.
\end{equation}
It corresponds to the partial wave expansion (\ref{partial}) with
$v_J=1$ and $v_J=0$ for $J\neq 1$. The angular dependence of the
pairing force in ${\rm He}^3$ is of this type. The gap equation
reads
\begin{equation}
\phi(\hat p)=\frac{3\lambda D_F}{8\pi}\int d^2\hat p^\prime\hat
p\cdot\hat p^\prime \phi(\hat
p^\prime)\ln\frac{2\omega_D}{|\phi(\hat p^\prime)|}
\label{gapeqlog}
\end{equation}
and its solution of zero azimuthal quantum number is given by
\begin{equation}
\phi(\hat p)=\phi_0\cos\theta=\phi_0P_1(\cos\theta)
\end{equation}
with $\theta$ the angle with respect to a prefixed direction in space and
\begin{equation}
\phi_0=2\omega_D e^{-\frac{2}{\lambda D_F}+\frac{1}{3}}.
\label{polarHe}
\end{equation}
The condensation energy density
\begin{equation}
F=-\frac{\omega_D^2D_F}{3}e^{-\frac{4}{\lambda D_F}+\frac{2}{3}}
\simeq -0.6492\omega_D^2D_Fe^{-\frac{4}{\lambda D_F}}.
\label{condI}
\end{equation}
This solution corresponds to the polar phase of ${\rm He}^3$
\cite{mahan}. Since the gap function contains only the partial
wave of $J=1$, there is no angular momentum mixing. The additional
term in the exponent of (\ref{polarHe}), $\frac{1}{3}$, comes from
the logarithm of (\ref{gapeqlog}).

\noindent
{\it {Case II}}:
\begin{equation}
V(\hat p\cdot\hat p^\prime)=4\pi\delta^2(\hat p-\hat p^\prime)
=\sum_J(2J+1)P_J(\hat p\cdot\hat p^\prime). \label{toy2}
\end{equation}
This corresponds to a singularity of the two body scattering
amplitudes in the forward direction. We have $v_J=1$ for all $J$
in (\ref{partial}). The last step of (\ref{toy2}) follows from the
addition theorem and the completeness of the spherical harmonics.
The gap equation (\ref{gapeqwk}) becomes
\begin{equation}
1=\frac{\lambda D_F}{2}\ln\frac{2\omega_D}{|\phi(\hat p)|},
\end{equation}
which implies a constant $|\phi(\hat p)|$ and yields a solution of odd parity
and zero azimuthal quantum number.
\begin{equation}
\phi(\hat p)=2\omega_De^{-\frac{2}{\lambda D_F}}{\rm
sign}(\cos\theta) =2\omega_De^{-\frac{2}{\lambda
D_F}}\sum_{n=0}^\infty (-1)^n(4n+3)
\frac{(2n-1)!!}{2^{n+1}(n+1)!}P_{2n+1}(\cos\theta).
\label{toymixing}
\end{equation}
The condensation energy density in this case reads
\begin{equation}
F=-\omega_D^2D_Fe^{-\frac{4}{\lambda D_F}}.
\end{equation}
We refer to this case as the case with the angular momentum mixing
because the gap function (\ref{toymixing}) contains all partial
waves. Carrying the solution of the case I to the case II amounts
to drop all partial waves other than that of $J=1$ and would lead
to a lower magnitude of the condensation energy (\ref{condI}).

The case with QCD is similar to the case II above since the
forward singularity of the diquark scattering renders the pairing
strength of all partial waves equal to the leading order.
The running coupling constant $g$ of QCD corresponds to $\lambda$ here
and the angular momentum mixing shows up in the $O(1)$ term of
$\ln|\phi|$. Therefore we expect angular momentum mixing to the subleading order
of the angular dependence of the gap function. Besides being an
ultra relativistic system, the CSC of QCD differs from the toy
model considered above in two aspects. The forward singularity of QCD also
brings about the energy dependence of the gap,
so the gap equation (\ref{gapeq}) will be replaced by the
Eliashberg equation derived by minimizing the CJT
effective action of QCD. Secondly, the pairing strength of
each partial wave does fall off with an increasing $J$ in the
sub-leading order of the pairing potential. It is this falling off that
makes the amount of the angular momentum mixing numerically small for
the solution considered in this article.

\section{Derivation of the gap equation from the CJT free energy}

\begin{figure}
\includegraphics[scale=0.6, clip=true]{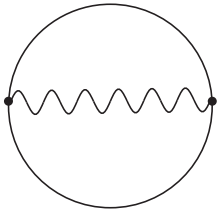}% Here is how to import EPS art
\caption{\label{fig:sset} The sunset diagram of Eq.(\ref{sun}).
Straight line and wavy line denotes quark and gluon propagators
respectively.}
\end{figure}

The QCD Lagrangian for one flavor of massless quark is given by
\begin{equation}
{\cal L}={\bar \psi}(i\gamma^{\mu}D_{\mu}+
\mu\gamma_0)\psi-\frac{1}{4}G_{a}^{\mu\nu}G_{\mu\nu}^{a}
+\hbox{renormalization counterterms}
\label{QCDL}
\end{equation}
where, $\psi$ is the quark spinor in Dirac and color space and
$\bar \psi=\psi^{\dagger}\gamma_0$. The covariant derivative
acting on the fermion field is $D_{\mu}=\partial_{\mu}+igT_a
A_{\mu}^{a}$, where $g$ is the running coupling constant,
$A_{\mu}^{a}$ is the gauge potential, $T_a=\lambda_a/2(a=1,...,8)$
is the $a$-th $SU(3)_c$ generator with $\lambda_a$ the $a$-th Gell-Mann
matrix. $G_{\mu\nu}^{a}=\partial_\mu A_{\nu}^{a}-\partial_\nu
A_{\mu}^{a}+gf^{abc}A_{\mu}^{b}A_{\nu}^{c}$ is the field strength
tensor. Introducing the Nambu-Gorkov spinor
\begin{equation}
\Psi= \left( \begin{array}{c} \psi\\
\psi_C
\end{array}
\right), \hspace{0.3cm}\bar \Psi=(\bar \psi, \bar \psi_C)
\label{NGspace}
\end{equation}
where $\psi_C=C{\bar \psi}^{T}$ is the charge-conjugate spinor and
$C\equiv i\gamma^2\gamma^0$, the CJT effective action
reads\cite{dk,CJT}
\begin{equation}
\Gamma[D,S]=-\frac{1}{2} \{{\rm Trln}D^{-1}+{\rm
Tr}(D_{0}^{-1}D-1)-{\rm Trln}S^{-1}-{\rm
Tr}(S_{0}^{-1}S-1)-2\Gamma_2[D,S]\} \label{2}
\end{equation}
where $D$ and $S$ are the full gluon and quark propagators,
$D_{0}^{-1}$ and $S_{0}^{-1}$ are the inverse tree-level
propagators for gluons and quarks, respectively. $\Gamma_2$ is the
sum of all two-particle irreducible(2PI)vacuum diagrams built with
$D$, $S$ and the tree-level quark-gluon vertex $\hat\Gamma$. We have
\begin{equation}
\Gamma_2=-\frac{1}{4}{\rm Tr}(D\hat{\Gamma}S\hat{\Gamma}S)+...,
\label{sun}
\end{equation}
where the first term corresponds to the sunset diagram of
Fig.\ref{fig:sset} and the contribution from ... is beyond the
subleading order of the gap function\cite{dk} .

The stationary points of the CJT effective action are determined
by
\begin{equation}
\frac{\delta\Gamma}{\delta D}=0,\hspace{0.5cm}
\frac{\delta\Gamma}{\delta S}=0
\label{3}
\end{equation}
which gives rise to the Dyson-Schwinger equation for gluons and
quarks,
\begin{equation}
\Pi\equiv-2\frac{\delta\Gamma_2}{\delta D},\hspace{0.5cm}
\Sigma\equiv2\frac{\delta\Gamma_2}{\delta S}
\label{5}
\end{equation}
where $\Pi$ and $\Sigma$ are the gluon and quark self-energy
defined via $D^{-1}=D_{0}^{-1}+\Pi$ and
$S^{-1}=S_{0}^{-1}+\Sigma$. Instead of solving the two equations
of (\ref{5}) simultaneously, we shall reduce the CJT effective
free energy with the aid of the first equation, leaving the gap
function arbitrary. The gap equation ( which is the Nambu-Gorkov
off diagonal part of the second equation of (\ref{5}) ) will be
derived after the CJT free energy is fully simplified under the
weak coupling approximation.

Substituting the first equation of (\ref{5}) into (\ref{2}),
the second term in Eq. ({\ref{2}}) cancels the last term. We have
\begin{equation}
\Gamma[D,S]=-\frac{1}{2}[{\rm Trln}D^{-1}-{\rm Trln}S^{-1}-{\rm
Tr}(S_{0}^{-1}S-1)] \label{6}
\end{equation}
In Nambu-Gorkov space, the inverse free quark propagator is
\begin{equation}
S_{0}^{-1}\equiv \left( \begin{array}{cc}
   {[G_0^ +  ]^{ - 1} } & 0  \\
   0 & {[G_0^ -  ]^{ - 1} }
\end{array} \right)
\label{7}
\end{equation}
where
\begin{equation}
[G_{0}^{\pm}]^{-1}=(p_0\pm\mu)\gamma_0-\vec{\gamma}\cdot\vec{p}.
\label{8}
\end{equation}
On writing the quark self-energy
\begin{equation}
\Sigma\equiv \left( \begin{array}{cc}
\Sigma^{+}&\Phi^{-}\\
\Phi^{+}&\Sigma^{-}
\end{array}\right)
\label{9}
\end{equation}
the full quark propagator,
\begin{equation}
S= \left( \begin{array}{cc}
   G^{+}  & \Xi^{-}  \\
   \Xi^{+} & G^{ - }
\end{array} \right)
\label{10}
\end{equation}
can be obtained explicitly by inverting the matrix $S_0^{-1}+\Sigma$.

For the single flavor pairing, the simplest choice of the off-diagonal
block of Eq. (\ref{9}) reads
\begin{equation}
\Phi^{+}(P)=i\phi\gamma_5\lambda_2
\label{13}
\end{equation}
and $\Phi^{+}=\Phi^{-}$(see Theorem 2 in\cite{BDJH}), where
$\lambda_2$ is the 2nd Gell-Mann matrix and $\phi$ is a function of
the energy and the momentum, i.e.
$\phi=\phi(p_0,\vec p)$. $\phi$ is even in $p_0$ and odd in
$\vec p$. By using the energy projectors of massless fermions
$\Lambda_p^{\pm}=(1\pm \gamma_0{\vec \gamma}\cdot{\hat p})/2$ and
ignoring the contribution from the wave-function renormalization,
the NG blocks of the propagator (\ref{10}) take the form
\begin{equation}
G^{\pm}=\frac{p_0+(p\mp\mu)}{p_0^2-(p\mp\mu)^2-\phi^2\lambda_2^2}\Lambda_p^+\gamma_0
+\frac{p_0(p\pm\mu)}{p_0^2-(p\pm\mu)^2-\phi^2\lambda_2^2}\Lambda_p^-\gamma_0
\label{14}
\end{equation}
\begin{equation}
\Xi^{\pm}=\frac{i\phi\lambda_2\gamma_5}{p_0^2-(p\pm\mu)^2-\phi^2\lambda_2^2}\Lambda_p^+
+\frac{i\phi\lambda_2\gamma_5}{p_0^2-(p\mp\mu)^2-\phi^2\lambda_2^2}\Lambda_p^-.
\label{15}
\end{equation}
Because of the $\lambda_2$ of (\ref{13}), the excitation in the third color direction 
is ungapped.

Now, we proceed to simplify the CJT free energy under the weak
coupling approximation. Denote by $\Gamma_n$ the free energy
density of the normal phase, we have
\begin{equation}
\Gamma = \Gamma_n + \Omega F
\end{equation}
where the condensate energy density
\begin{equation}
F=-\frac{1}{2\Omega}[{\rm Trln}D^{-1}-{\rm Trln}D_n^{-1}-{\rm Trln}S^{-1}
+{\rm Trln}S_0^{-1}-{\rm Tr}(S_{0}^{-1}S-1)]
\end{equation}
is the part of $\Gamma$ responsible to the gap equation. Following the procedure of
\cite{IDHD,JHIDD}, we approximate
\begin{equation}
{\rm Trln}D^{-1}-{\rm Trln}[D_n^{-1}]\simeq{\rm Tr}[D_n\delta\Pi]
\label{16}
\end{equation}
where
\begin{equation}
\delta\Pi=\Pi-\Pi_n,
\end{equation}
with $\Pi_n$ the hard-dense-loop (HDL)
resummed gluon self-energy in normal phase and $D_n$ the corresponding HDL
gluon propagator.
In the Coulomb gauge, the HDL gluon propagator is
\begin{equation}
D_{n,00}(K)=D_l(K), \hspace{0.3cm}
D_{n,0i}(K)=D_{n,i0}=0,\hspace{0.3cm}
D_{n,ij}=(\delta_{ij}-\hat{k}_i\hat{k}_j)D_t(K) \label{18}
\end{equation}
where $D_{l,t}$ are the longitudinal and transverse propagators
respectively and are diagonal in adjoint color space, i.e.
$D_{l,t}^{ab}=\delta^{ab} D_{l,t}$. Consequently, we only need the
00-component, $\Pi^{00}(K)$, and the transverse projection of the
ij-components,
\begin{equation}
(\delta_{ij}-\hat{k}_i\hat{k}_j)\Pi^{ij}(K)=\Pi^{ii}(K)-\hat{k}_i\hat{k}_j\Pi^{ij}(K)
\label{19}
\end{equation}
The gluon self-energy in super phase reads
\begin{equation}
\Pi_{ab}^{\mu\nu}(K)=\frac{1}{2}T\sum\limits_{P,P^{\prime}}{\rm
Tr}[\hat{\Gamma}_{a}^{\mu}S(P)\hat{\Gamma}_{b}^{\mu}S(P^{\prime})]
\label{20}
\end{equation}
where $K=P-P^{\prime}$ and
\begin{equation}
\hat\Gamma_{\mu}^{a}\equiv \left (
\begin{array}{cc}
\Gamma_{\mu}^{a} & 0 \\
0 & \bar\Gamma_{\mu}^{a}
\end{array} \right )
\label{21}
\end{equation}
with $\Gamma_{a}^{\mu}=\gamma^{\mu}T_a$ and
$\bar\Gamma_{a}^{\mu}=-\gamma^{\mu}T_a^{T}$. Substituting Eq. (\ref{10})
into Eq.(\ref{20}), we find that
Nambu-Gorkov space,
\begin{eqnarray}
\nonumber \Pi_{ab}^{\mu\nu}(K)&=&
\frac{1}{2}T\sum\limits_{P,P^{\prime}}\big\{\rm
{Tr}[\Gamma_{a}^{\mu}G^+(P)\Gamma_{b}^{\nu}G^+(P^{\prime})]
+{\rm Tr}[\bar\Gamma_{a}^{\mu}G^-(P)\bar\Gamma_{b}^{\nu}G^-(P^{\prime})]\\
&&+{\rm
Tr}[\Gamma_{a}^{\mu}\Xi^-(P)\bar\Gamma_{b}^{\nu}\Xi^+(P^{\prime})]
+{\rm
Tr}[\bar\Gamma_{a}^{\mu}\Xi^+(P)\Gamma_{b}^{\nu}\Xi^-(P^{\prime})]\big\}
\label{22}
\end{eqnarray}
Since the HDL gluon propagators are diagonal in color space, we
only need the diagonal terms of Eq.(\ref{22}) to deal with Eq.(\ref{16}).
The explicit form of each diagonal term of (\ref{22}) reads
\begin{subequations}
\begin{eqnarray}
&&{\rm
Tr}[\Gamma_{a}^{\mu}G^+(P)\Gamma_{a}^{\nu}G^+(P^{\prime})]=\frac{g^2}{2}T\sum\limits_{P,P^{\prime}}
{\rm
Tr}\big[\gamma^\mu\Lambda_{p}^{+}\gamma_0\gamma^{\nu}\Lambda_{p^{\prime}}^{+}\gamma_0
\big]w_a^+(P,P^{\prime}),\label{23a}
\\
&&{\rm
Tr}[\bar\Gamma_{a}^{\mu}G^-(P)\bar\Gamma_{a}^{\nu}G^-(P^{\prime})]=\frac{g^2}{2}T\sum\limits_{P,P^{\prime}}
{\rm
Tr}\big[\gamma^\mu\Lambda_{p}^{-}\gamma_0\gamma^{\nu}\Lambda_{p^{\prime}}^{-}\gamma_0
\big]w_a^-(P,P^{\prime}),\label{23b}
\\
&&{\rm
Tr}[\Gamma_{a}^{\mu}\Xi^-(P)\bar\Gamma_{a}^{\nu}\Xi^+(P^{\prime})]=-\frac{g^2}{2}T\sum\limits_{P,P^{\prime}}
{\rm
Tr}\big[\gamma^\mu\gamma_5\Lambda_{p}^{+}\gamma^{\nu}\gamma_5\Lambda_{p^{\prime}}^{-}
\big]w_a(P,P^{\prime}),\label{23c}
\\
&&{\rm
Tr}[\bar\Gamma_{a}^{\mu}\Xi^+(P)\Gamma_{a}^{\nu}\Xi^-(P^{\prime})]=-\frac{g^2}{2}T\sum\limits_{P,P^{\prime}}
{\rm
Tr}\big[\gamma^\mu\gamma_5\Lambda_{p}^{-}\gamma^{\nu}\gamma_5\Lambda_{p^{\prime}}^{+}
\big]w_a(P,P^{\prime}).\label{23d}
\end{eqnarray}
\end{subequations}
where the repeated color indexes on LHS are {\it not} to be
summed. The quantities $w^{\pm}$ and $w$ on RHS of
Eqs.(\ref{23a}-\ref{23d}) are given by
\begin{subequations}
\begin{equation}
w_a^{\pm}=\left\{
\begin{array}{ll}
\begin{gathered}
\frac{1}{2}\frac{p_0\pm\epsilon_p}{p_{0}^{2}-\varepsilon_{p}^{2}}\frac{p_0^{\prime}\pm\epsilon_{p^{\prime}}}{p_{0}^{\prime2}-\varepsilon_{p^{\prime}}^{2}}
\hspace{3.3cm},  a =1,2,3
\end{gathered}\\
\hfill \\
\begin{gathered}
\frac{1}{4}\Big[\frac{p_0\pm\epsilon_p}{p_{0}^{2}-\varepsilon_{p}^{2}}\frac{p_0^{\prime}\pm\epsilon_{p^{\prime}}}{p_{0}^{\prime2}-\epsilon_{p^{\prime}}^{2}}
+\frac{p_0\pm\epsilon_p}{p_{0}^{2}-\epsilon_{p}^{2}}\frac{p_0^{\prime}\pm\epsilon_{p^{\prime}}}{p_{0}^{\prime2}-\varepsilon_{p^{\prime}}^{2}}\Big],
a=4,\cdot\cdot\cdot,7
\end{gathered}\\
\hfill\\
\begin{gathered}
\frac{1}{6}\frac{p_0\pm\epsilon_p}{p_{0}^{2}-\varepsilon_{p}^{2}}\frac{p_0^{\prime}\pm\epsilon_{p^{\prime}}}{p_{0}^{\prime2}-\varepsilon_{p^{\prime}}^{2}}
+\frac{1}{3}\frac{p_0\pm\epsilon_p}{p_{0}^{2}-\epsilon_{p}^{2}}\frac{p_0^{\prime}\pm\epsilon_{p^{\prime}}}{p_{0}^{\prime2}-\epsilon_{p^{\prime}}^{2}}
\hspace{0.1cm},a=8
\end{gathered}
\end{array}
\right. \label{24a}
\end{equation}
and
\begin{equation}
w_a=\left\{
\begin{array}{ll}
\begin{gathered}
-\frac{1}{2}\frac{\phi(P)\phi(P^{\prime})}{(p_{0}^{2}-\varepsilon_{p}^{2})(p_{0}^{\prime
2}-\varepsilon_{p^{\prime}}^{2})}\hspace{0.1cm},  a=1,2,3
\end{gathered}\\
\hfill\\
\begin{gathered}
0\hspace{3.5cm},a=4,\cdot\cdot\cdot,7
\end{gathered}\\
\hfill\\
\begin{gathered}
\frac{1}{6}\frac{\phi(P)\phi(P^{\prime})}{(p_{0}^{2}-\varepsilon_{p}^{2})(p_{0}^{\prime
2}-\varepsilon_{p^{\prime}}^{2})}\hspace{0.4cm},  a=8
\end{gathered}
\end{array}
\right. \label{24b}
\end{equation}
\end{subequations}
where $\epsilon_p=p-\mu$ and
$\varepsilon_p=\sqrt{(p-\mu)^2+\phi^2(P)}$. Since the dominant
contributions in the weak coupling arise from the quasiparticles,
we have ignored the contributions from the quasi-antiparticles in
the calculations above. The trace over Dirac space is
straightforward
\begin{subequations}
\begin{eqnarray}
&&{\rm
Tr}\big[\gamma^{0}\Lambda_{p}^{\pm}\gamma_0\gamma^{0}\Lambda_{p^{\prime}}^{\pm}\gamma_0\big]
=-{\rm Tr}\big[\gamma^0\gamma_5\Lambda_{p}^{\pm}\gamma^{0}\gamma_5\Lambda_{p^{\prime}}^{\mp}\big]=1+\hat{p}\cdot\hat{p^{\prime}},
\label{25a}\\
\hfill\nonumber\\
&&\sum\limits_{i}{\rm
Tr}\big[\gamma^{i}\Lambda_{p}^{\pm}\gamma_0\gamma^{i}\Lambda_{p^{\prime}}^{\pm}\gamma_0\big]
=\sum\limits_{i}{\rm
Tr}\big[\gamma^{i}\gamma_5\Lambda_{p}^{\pm}\gamma^i\gamma_5\Lambda_{p^{\prime}}^{\mp}\big]=3-\hat{p}\cdot\hat{p^{\prime}},
\label{25b}\\
&&{\rm
Tr}\big[\vec{\gamma}\cdot\vec{k}\Lambda_{p}^{\pm}\gamma_0\vec{\gamma}\cdot\vec{k}\Lambda_{p^{\prime}}^{\pm}\gamma_0\big]
={\rm
Tr}\big[\vec{\gamma}\cdot\vec{k}\gamma_5\Lambda_{p}^{\pm}\vec{\gamma}\cdot\vec{k}\gamma_5\Lambda_{p^{\prime}}^{\mp}\big]
=(1+\hat{p}\cdot\hat{p^{\prime}})\frac{(p-p^{\prime})^2}{k^2}
\label{25c}
\end{eqnarray}
\end{subequations}

It can be shown that the contribution from Eq.(\ref{24a}) to $F$
is suppressed by an order $g$ relative to that from Eq.(\ref{24b})
and will be ignored here. We neglect also the dependence of the
gap function on the magnitude of the momentum, but keep the
dependence on the energy and the momentum orientation. Then
the integrals over $p$ and $p^{\prime}$ can be carried out easily. Since we are only
interested in the zero temperature, the Matsubara sum becomes an
integral over the Euclidean energy. We find
\begin{equation}
{\rm Tr}[D_n\delta\Pi]=-\frac{6\bar{g}^2\mu^4}{32\pi^4}\int d\nu \int
d\nu^{\prime}\int d^2\hat{p}\int
d^2\hat{p^{\prime}}\frac{\phi(\nu,\hat{p})\phi(\nu^{\prime},\hat{p^{\prime}})}{\sqrt{(\nu^2+\phi^2(\nu,\hat{p}))(\nu^{\prime2}+\phi^2(\nu^{\prime},\hat{p^{\prime}})})}
\times\big[D_l(\nu-\nu^{\prime},\theta)+D_t(\nu-\nu^{\prime},\theta)\big]
\label{26}
\end{equation}
where $\cos\theta=\hat p\cdot\hat p^\prime$. Making use of the Nambu-Gorkov formalism in
Eq.(\ref{7}-\ref{15}), the rest terms of the condensate
energy density Eq.(\ref{6}) can be evaluated readily.
\begin{equation}
\frac{1}{\Omega}{\rm Trln}S^{-1}-{\rm Trln}S_0^{-1}=-\frac{4\mu^2}{(2\pi)^3}\int d\nu\int
d^2\hat{p}\Big[|\nu|-\sqrt{\nu^2+\phi^2(\nu,\hat{p})}\Big]
\label{27}
\end{equation}
\begin{equation}
\frac{1}{\Omega}{\rm
Tr}(S_{0}^{-1}S-1)=-\frac{4\mu^2}{(2\pi)^3}\int d\nu\int
d^2\hat{p}\frac{\phi^2(\nu,\hat{p})}{\sqrt{\nu^2+\phi^2(\nu,\hat{p})}}
\label{28}
\end{equation}
The final expression of the condensation energy density reads
\begin{eqnarray}
\nonumber
F=&-&\frac{3\bar{g}^2\mu^4}{32\pi^4}\int d\nu \int
d\nu^{\prime}\int d^2\hat{p}\int
d^2\hat{p^{\prime}}V(\nu-\nu^{\prime},\theta)\frac{\phi(\nu,\hat{p})\phi(\nu^{\prime},\hat{p^{\prime}})}{\sqrt{(\nu^2+\phi^2(\nu,\hat{p}))(\nu^{\prime2}+\phi^2(\nu^{\prime},\hat{p^{\prime}})})}\\
&+&\frac{2\mu^2}{(2\pi)^3}\int d\nu\int
d^2\hat{p}\frac{\nu^2}{\sqrt{\nu^2+\phi^2(\nu,\hat{p})}}
\label{29}
\end{eqnarray}
where $V$ contains the contribution from both magnetic and
electric gluons, i.e.
\begin{equation}
V=D_l(\nu-\nu^{\prime},\theta)+D_t(\nu-\nu^{\prime},\theta)
\label{30}
\end{equation}
The gap equation can be derived by minimizing $F$
with respect to the gap function $\phi(\nu,\hat{p})$,
\begin{equation}
\frac{\delta\Gamma}{\delta\phi}=0
\label{31}
\end{equation}
and we end up with
\begin{equation}
\phi(\nu,\hat{p})=\frac{g^2\mu^2}{24\pi^3}\int d\nu^{\prime}\int
d^2\hat{p}^{\prime}V(\nu-\nu^{\prime},\theta)\frac{\phi(\nu^{\prime},\hat{p}^{\prime})}{\sqrt{\nu^{\prime2}+\phi^2(\nu^{\prime},\hat{p}^{\prime})}}
\label{qcdgapeq}
\end{equation}

A consistent derivation of the gap equation up to the subleading
order requires both the contribution from $w_a^{\pm}$ and that
from the diagonal block of (\ref{9}) to be kept. The net result is
to replace the first term inside the square root on RHS of
(\ref{qcdgapeq}) by $\nu^{\prime 2}/Z^2(\nu^\prime)$ with $Z(\nu)$
the wave function renormalization of the normal phase. But it will
not interfere with the angular dependence of the gap function to
the subleading order as will be shown in the next section.

\section{The integral equation for the angular dependence of the gap}

Although the pairing strength are equal to the leading order of
the QCD running coupling constant, similar to the case II of the
toy model, the subleading terms fall off with an increasing $J$.
This makes the solution to the gap equation (\ref{qcdgapeq})
highly nontrivial. In what follows, we shall isolate the energy
dependence and the angle dependence of the pairing potential $V$.
A differential equation with respect to the Matsubara energy will
be derived from (\ref{qcdgapeq}) that fixes the gap function up to
an arbitrary function of the angle. This function will be
determined then by (\ref{qcdgapeq}) with $\phi$ a known function
of the Matsubara energy.

Proceeding with the partial wave analysis, we
expand $V(\nu-\nu^{\prime},\theta)$ in series of Legendre
polynomials\cite{WJH}:
\begin{equation}
V(\nu-\nu^{\prime},\theta)=\frac{1}{6\mu^2}{\rm
ln}\frac{\omega_c}{|\nu-\nu^{\prime}|}\sum\limits_{J}(2J+1)P_J(\cos\theta)+\frac{1}{2\mu^2}\sum\limits_{J>0}
(2J+1)c_JP_J(\cos\theta)
\label{33}
\end{equation}
where $\omega_c=\frac{1024\sqrt{2}\pi^4\mu}{N_f^{\frac{5}{2}}g^5}$ and $c_J$ is
given by Eq.(\ref{subleading}).
Using the completeness relation
\begin{equation}
\sum\limits_{J}(2J+1)P_J(\cos\theta)=4\pi\delta^2(\hat{p}-\hat{p}^{\prime})
\label{35}
\end{equation}
and the identity (proved in the Appendix A)
\begin{equation}
\int
d^2\hat{p}^{\prime}\sum\limits_{J=1}^{\infty}(2J+1)c_JP_J(\hat{p}\cdot\hat{p}^{\prime})f(\hat{p}^{\prime})
=2\int
d^2\hat{p}^{\prime}\frac{f(\hat{p}^{\prime})-f(\hat{p})}{|1-\hat{p}\cdot\hat{p}^{\prime}|}
\label{36}
\end{equation}
with $f(\hat{p})$ an arbitrary function of $\hat{p}$, the gap equation (\ref{qcdgapeq}) becomes
\begin{eqnarray}
\nonumber
\phi(\nu,x)=&&\bar{g}^2\int\limits_{0}^{\omega_0}d\nu^{\prime}\Big\{\frac{1}{2}\big({\rm
ln}\frac{\omega_c}{|\nu-\nu^{\prime}|}+{\rm
ln}\frac{\omega_c}{|\nu+\nu^{\prime}|}\big)\frac{\phi(\nu^{\prime},x)}{\sqrt{\nu^{\prime
2}+\phi^2(\nu^{\prime},x)}}\\
&&+3\int\limits_{-1}^{1}dx^{\prime}\frac{1}{|x-x^{\prime}|}\Big[\frac{\phi(\nu^{\prime},x^{\prime})}
{\sqrt{\nu^{\prime2}+\phi^2(\nu^{\prime},x^{\prime})}}-\frac{\phi(\nu^{\prime},x)}
{\sqrt{\nu^{\prime2}+\phi^2(\nu^{\prime},x)}}\Big]\Big\}
\label{37}
\end{eqnarray}
where $\bar g^2=g^2/(18\pi^2)$, $x=\hat p\cdot\hat z$ with $\hat z$ a fixed spatial direction
and a UV cutoff, $\omega_0\sim g\mu$ is introduced. In deriving (\ref{37}),
we have assumed that the gap depends on $x$ only, so the integration over the
azimuthal angle of $\hat p^\prime$ can be carried out explicitly.
The gap equation (\ref{37}) can be
further simplified by using the approximation of Son \cite{D}
\begin{equation}
{\rm ln}\frac{\omega_c}{|\nu-\nu^{\prime}|}\simeq{\rm
ln}\frac{\omega_c}{|\nu_>|}
\label{son}
\end{equation}
with $\nu_>={\rm max}(\nu,\nu^\prime)$. It is convenient to
introduce
\begin{equation}
\xi=\rm ln \frac{\omega_c}{\nu},\hspace{0.5cm}
a=\rm ln\frac{\omega_c}{\omega_0}.
\label{newvar}
\end{equation}
On writing $\phi=\phi(\xi,x)$ and
\begin{equation}
\Phi(\xi,x)\equiv\bar{g}^2\int\limits_{\xi}^{\infty}d\xi^{\prime}\frac{\phi(\xi^{\prime},x)}
{\sqrt{1+\frac{\phi^2(\xi^{\prime},x)}{\omega_c^2}e^{2\xi^{\prime}}}}
\label{38}
\end{equation}
the gap equation(\ref{37}) becomes
\begin{equation}
\phi(\xi,x)=\xi\Phi(\xi,x)-\int\limits_{a}^{\xi}d\xi^{\prime}\xi^{\prime}\frac{d\Phi}{d\xi^{\prime}}
+3\int\limits_{-1}^{1}dx^{\prime}\frac{\Phi(a,x^{\prime})-\Phi(a,x)}{|x-x^{\prime}|}
\label{39}
\end{equation}
Taking the derivative of both sides with respect to $\xi$, we find
\begin{equation}
\frac{d\phi}{d\xi}=\Phi(\xi,x)
\label{40}
\end{equation}
which implies the boundary condition
\begin{equation}
\frac{d\phi}{d\xi}\rightarrow 0
\label{41}
\end{equation}
as $\xi\rightarrow \infty$ for all $x$. Another derivative of
(\ref{40}) yield the ordinary differential equation
\begin{equation}
\frac{d^2\phi}{d\xi^2}+\frac{\bar{g}^2\phi}{\sqrt{1+\frac{\phi^2}{\omega_c^2}e^{2\xi}}}=0
\label{42}
\end{equation}
which is universal for all $x$. It follows from Eq.(\ref{39}) that
the gap equation is equivalent to
\begin{equation}
a\Phi(a,x)-\phi(a,x)+3\int\limits_{-1}^{1}\frac{\Phi(a,x^{\prime})-\Phi(a,x)}{|x-x^{\prime}|}=0
\label{43}
\end{equation}
The solution to (\ref{42}) subject to the condition (\ref{41}) contains an arbitrary function
of $x$ to be determined by (\ref{43}). No further approximation has been made up to now.

The solution to the differential equation (\ref{42}) proceeds in the same way as that for
a spherical gap. To the leading order, the equation can be
approximated by a linear one,
\begin{equation}
\frac{d^2\phi^{(0)}}{d\xi^2}+\bar{g}^2\theta(b-\xi)\phi^{(0)}=0
\label{44}
\end{equation}
where $b(x)$ is to be determined by the condition
$\frac{\phi(b,x)}{\omega_c}e^b=1$. Its solution that satisfies the
boundary condition(\ref{41}) and the continuity up to the first order
derivative reads
\begin{equation}
\phi^{(0)}(\xi,x) = \left\{
\begin{array}{ll}\phi_0(x)\cos\bar{g}[b(x)-\xi],\hspace{0.3cm}  {
\rm for}\hspace{0.2cm}
\xi<b(x),\\
\hfill \\
\phi_0(x),\hspace{2.5cm}  {\rm for}\hspace{0.2cm} \xi\ge b(x).
\end{array} \right. \label{45}
\end{equation}
where
\begin{equation}
b(x)={\rm ln}\frac{\omega_c}{|\phi_0(x)|}
\label{46}
\end{equation}
It follows from Eq.(\ref{40}) then that
\begin{equation}
\Phi(\xi,x) = \left\{
\begin{array}{ll}\bar{g}\phi_0(x)\sin\bar{g}[b(x)-\xi],\hspace{0.3cm}  {
\rm for}\hspace{0.2cm}
\xi<b(x),\\
\hfill \\
0,\hspace{3.3cm}  {\rm for}\hspace{0.2cm} \xi\ge b(x).
\end{array} \right.
\label{47}
\end{equation}
The angle dependent factor $f(\hat p)$ introduced in
Eq.(\ref{gap2}) is defined by
\begin{equation}
f(x)\equiv\frac{\phi_0(x)}{\Delta_0}=O(1)
\end{equation}
where $\Delta_0$ is the $s$-wave gap given by
\begin{equation}
\frac{\pi}{2}-\bar{g}{\rm ln}\frac{2\omega_c}{\Delta_0}=0.
\label{55}
\end{equation}
where the contribution from the wave-function renormalization is ignored.
Up to the subleading order, the differential equation(\ref{42}) reads
\begin{equation}
\frac{d^2\phi^{(1)}}{d\xi^2}+\bar{g}^2\theta(b-\xi)\phi^{(1)}=\bar{g}^2\Big[\theta(b-\xi)-\frac{1}
{\sqrt{1+\frac{\phi^2}{\omega_c^2}e^{2\xi}}}\Big]\phi^{(0)}. \label{48}
\end{equation}
We find that
\begin{equation}
\phi^{(1)}(\xi,x)=\phi^{(1)}(\xi,x)+A(\xi,x)u(\xi,x)-B(\xi,x)v(\xi,x),
\label{49}
\end{equation}
where $u(\xi,x)$ and $v(\xi,x)$ are the two linearly independent
solutions to the Eq.(\ref{44}),

\begin{equation}
u(\xi,x) = \left\{
\begin{array}{ll}\cos\bar{g}[b(x)-\xi],\hspace{0.3cm}  {
\rm for}\hspace{0.2cm}
\xi<b(x),\\
\hfill \\
1,\hspace{2.3cm}  {\rm for}\hspace{0.2cm} \xi\ge b(x).
\end{array}
\right.
\end{equation}

\begin{equation}
v(\xi,x) = \left\{
\begin{array}{ll}-\sin\bar{g}[b(x)-\xi],\hspace{0.3cm}  {
\rm for}\hspace{0.2cm}
\xi<b(x),\\
\hfill \\
\bar g\xi,\hspace{2.4cm}  {\rm for}\hspace{0.2cm} \xi\ge b(x).
\end{array}\right.
\end{equation}
and
\begin{subequations}
\begin{equation}
A(\xi,x)=\bar{g}\int\limits_{\xi}^{\infty}d\xi^{\prime}\Big[\theta(b-\xi^{\prime})-\frac{1}
{\sqrt{1+\frac{\phi^2(\xi^{\prime},x)}{\omega_c^2}e^{2\xi^{\prime}}}}\Big]v(\xi^{\prime},x)
\phi^{(0)}(\xi^\prime,x),
\label{50a}
\end{equation}
\begin{equation}
B(\xi,x)=-\bar{g}\int\limits_{\xi}^{\infty}d\xi^{\prime}\Big[\theta(b-\xi^{\prime})-\frac{1}
{\sqrt{1+\frac{\phi^2(\xi^{\prime},x)}{\omega_c^2}e^{2\xi^{\prime}}}}\Big]u(\xi^{\prime},x)
\phi^{(0)}(\xi^\prime,x).
\label{50b}
\end{equation}
\end{subequations}
At the point $\xi=a$, we have
\begin{equation}
A(a,x)=1+O(\bar{g}),\hspace{0.5cm} B(a,x)\simeq\bar{g}\rm ln2
\label{51}
\end{equation}
Therefore
\begin{equation}
\phi^{(1)}(a,x)\simeq\phi_0(x)\Big[\cos\bar{g}(b-a)-\bar{g}{\rm
ln}2\sin\bar{g}(b-a)\Big]
\label{52}
\end{equation}
to the subleading order.
Since $\bar{g}(b-a)=\pi/2+O(g)$ according to Eq.(\ref{55}), we have
\begin{equation}
\Phi(a,x)=\bar{g}\phi_0(x)+O(g),
\label{53_1}
\end{equation}
and
\begin{equation}
\phi^{(1)}(a,x)\simeq\phi_0(x)\Big[\frac{\pi}{2}-\bar{g}(b-a)-\bar{g}{\rm
ln}2\Big]+O(g)
\label{53}
\end{equation}
Substituting Eqs.(\ref{53_1}) and (\ref{53}) into Eq.(\ref{43}), we obtain the gap
equation to the subleading order
\begin{equation}
-\Big[\frac{\pi}{2}-\bar{g}{\rm
ln}\frac{\omega_c}{|\phi_0(x)|}-\bar{g}{\rm
ln}2\Big]\phi_0(x)+\int\limits_{-1}^{1}dx^{\prime}\frac{\phi_0(x^{\prime})-\phi_0(x)}{|x-x^{\prime}|}=0
\label{54}
\end{equation}
Then the integral equation for $f(x)$,
\begin{equation}
f(x){\rm
ln}|f(x)|-3\int\limits_{-1}^{1}dx^{\prime}\frac{f(x^{\prime})-f(x)}{|x-x^{\prime}|}=0
\label{56}
\end{equation}
follows from (\ref{55}).

Few comments are in order: 1) The spherical gap, $f(x)=1$ is a
trivial solution to Eq.(\ref{56}) and there is no angular momentum mixing.
2) The "spin-1" gap, carried over from the polar phase of ${\rm He^3}$, $f(x)\propto x$, fails
to satisfy this equation. 3) Eq.(\ref{56}) conserves the parity.
In another word, its solution can be either an even or an odd
function of $x$. 4) If the wavefunction renormalization is
restored, there will be an additional subleading term on RHS of
(\ref{48}) and an additional subleading term on RHS of $B$ of
eq.(\ref{51}). This term, when substitute into Eq.(\ref{43}), will
cancel the corresponding contribution to $\Delta_0$
\begin{equation}
\Delta_0=\pi e^{-\gamma}T_c^{(0)}=\frac{2048\pi^4\mu}{N_f^{\frac{5}{2}}g^5}
e^{-\frac{3\pi^2}{\sqrt{2}g}-\frac{\pi^2+4}{8}}.
\label{delta0}
\end{equation}
leaving the integral equation (\ref{56}) intact.

\section{The numerical results of the angular dependence}

The solution to the integral equation Eq.(\ref{56}) can be
obtained from a variational principle. Upon substitution of
Eq.(\ref{45}) with $\phi_0(x)=\Delta_0f(x)$ into Eq.(\ref{29}),
the condensate energy density becomes a functional of $f$(details in
Appendix B), i. e.
\begin{equation}
F=\frac{\mu^2\Delta_0^2}{2\pi^2}{\cal F}[f],
\end{equation}
where
\begin{eqnarray}
\nonumber {\cal F}[f(x)]&=&\int\limits_{-1}^{1}dxf^2(x)\Big[{\rm
ln}|f(x)|-\frac{1}{2}\Big]+\frac{3}{2}\int\limits_{-1}^{1}dx\int\limits_{-1}^{1}dx^{\prime}\frac{\big[f(x)-f(x^{\prime})\big]^2}
{|x-x^{\prime}|}\\
&=&2\int\limits_{0}^{1}dxf^2(x)\Big[{\rm
ln}|f(x)|-\frac{1}{2}\Big]+3\int\limits_{0}^{1}dx\int\limits_{0}^{1}dx^{\prime}\Big\{
\frac{\big[f(x)-f(x^{\prime})\big]^2}
{|x-x^{\prime}|}+\frac{\big[f(x)+f(x^{\prime})\big]^2}
{x+x^{\prime}}\Big\} \label{A11}
\label{functional}
\end{eqnarray}
with the last equality following from the odd parity of $f(x)$, i.
e. $f(-x)=-f(x)$. Readers may easily verify that the variational
minimum of Eq.(\ref{functional}) does solve Eq.(\ref{56}).

Before the numerical solution, we consider a trial function
\begin{equation}
f(x)=cx
\label{trial}
\end{equation}
and substitute it into the target functional
(\ref{functional}). The minimization yields
\begin{equation}
c=e^{-\frac{17}{3}}=e^{-6+\frac{1}{3}}\simeq 3.459\times 10^{-3},
\label{prefactor}
\end{equation}
at which
\begin{equation}
{\cal F}[f]\simeq -3.989\times 10^{-6}. \label{initial}
\end{equation}
The trial function (\ref{trial}) is what people carried over from
the polar phase of ${\rm He^3}$. The "-6" of the exponent of
(\ref{prefactor}) comes from the pairing strength of the $p$-wave
and the "$\frac{1}{3}$" stems from the logarithm of (\ref{56}). 
The latter contribution was reported in \cite{A}.
The trial function (\ref{trial})
with $p$-wave alone is not optimal. The free energy will be
lowered further by including higher partial waves of odd $J$ as we
shall see.

To find the variational minimum, we discretize the integral of
Eq.(\ref{functional}) by dividing the domain $x\in (0,1)$ into
$N(>>1)$ intervals with
\begin{equation}
x_j=(j+\frac{1}{2})\Delta x, \hspace{0.5cm}
j=0,1,2,\cdot\cdot\cdot,N-1
\end{equation}
where $\Delta x=1/N$. We have then ${\cal F}=\lim_{N\to\infty}{\cal F}_N$ with
\begin{equation}
{\cal F}_N= 2\Delta x\sum\limits_{j}f_{j}^{2}\Big({\rm
ln}f_j-\frac{1}{2}\Big)+6\Delta x^2\sum\limits_{j}\frac{f_j^2}{x_j}+3\Delta x^2\sum\limits_{i,j,i\neq
j}\Big[\frac{(f_i-f_j)^2}{|x_i-x_j|}+\frac{(f_i+f_j)^2}{x_i+x_j}\Big]
\label{A13}
\end{equation}
where we have dropped the limit $x^{\prime}\rightarrow x$ of the
first term inside the curly bracket of Eq. (\ref{functional}). ${\cal F}_N$ is a
function of $N$ variables. The stationary condition
\begin{equation}
\frac{\partial {\cal F}}{\partial f_j}=0
\label{A14}
\end{equation}
yields
\begin{equation}
f_j{\rm ln}f_j+3\Delta x\Big[\frac{2}{x_j}+\sum\limits_{i\neq
j}\Big(\frac{1}{|x_i-x_j|}+\frac{1}{x_i+x_j}\Big)\Big]f_j-3\Delta x\sum\limits_{i\neq
j}\Big(\frac{1}{|x_i-x_j|}+\frac{1}{x_i+x_j}\Big)f_i=0 \label{A15}
\end{equation}
which is a discrete version of Eq. (\ref{56}). Regarding $f_i$'s
as given, the equation for $f_j$ is of the form
\begin{equation}
({\rm ln}f_j+a)f_j-b=0
\label{A16}
\end{equation}
with $a$ and $b$ positive. It has one and only one solution for
$f_j>0$.

\begin{figure}
\includegraphics[scale=1.0, clip=true]{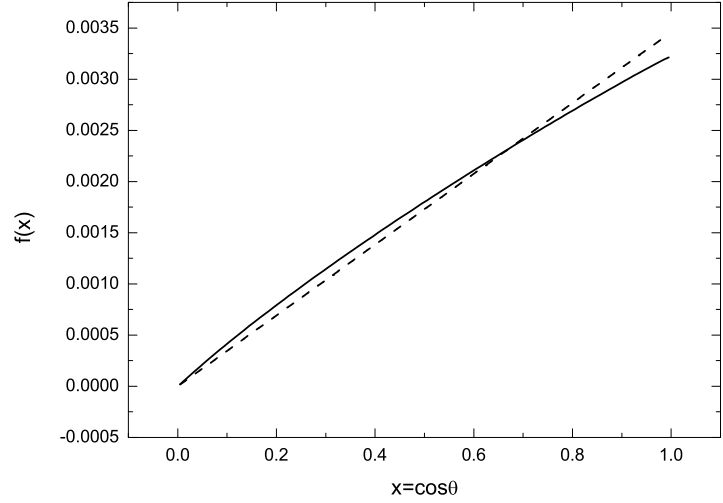}% Here is how to import EPS art
\caption{\label{fig:epsart} The angular dependence of the gap
function with angular momentum mixing. The dashed line and the
solid line are the initial configuration and the final numerical
results respectively.}
\end{figure}

We start with the trial function (\ref{trial}),
$f_j=e^{-\frac{17}{3}}x_j$ as an initial configuration and update
each $f_j$ by solving Eq.(\ref{A16}). This way we lower the value
of the target functional ${\cal F}$ in each step and approach the
solution to (\ref{56}) eventually. The process converges rapidly
and our numerical solution to (\ref{56}) is shown as solid line in
Fig.\ref{fig:epsart}, which depart from the trial function (dashed
line) slightly. We find the minimum value of the target functional
\begin{equation}
{\cal F}[f]\simeq -4.130\times 10^{-6}.
\end{equation}
which drops from (\ref{initial}) by $3.5$ percent.

It is instructive to examine the angular momentum contents of our solution in the
partial wave expansion
\begin{equation}
f(x) = \sum_{J={\rm odd}}b_JP_J(x).
\end{equation}
The coefficients of the first three partial waves, $J=1,3,5$, calculated by
substituting the numerical solution into the formula
\begin{equation}
b_J=\frac{2J+1}{2}\int\limits_{-1}^{1}dxf(x)P_J(x)
\end{equation}
are displayed in Table I. While
the gap function contains all partial waves of odd $J$, the component of $J=1$
is the biggest. This is anticipated because
the pairing strength of the all partial waves are equal in leading
order but fall off with an increasing $J$ in the subleading order
as is shown in the partial wave expansion (\ref{33}).

\begin{table}
\begin{tabular}{c|ccc}
\hline
\hline
\hfill\\
&\quad\quad$J=1\quad\quad$ &$J=3\quad\quad$ &$J=5$\\
\hfill\\
\hline \hfill\\$b_J$\quad\quad & $\quad\quad3.413\times 10^{-3}\quad\quad $&$-2.328\times 10^{-4}\quad\quad$ & $7.409\times 10^{-5}$\\
\hfill\\
\hline \hline
\end{tabular}
\caption{The first three expansion coefficients of the gap
function according to Legendre polynomials.}
\end{table}

\section{Concluding remarks}

%%%%%%%%%%%%%%%%%%%%%%%%%

In summary we have explored the angular dependence of the gap
function for a non-spherical pairing of CSC. Because of the equal
strength of the pairing potential mediated by one-gluon exchange
for all partial waves to the leading order of QCD running coupling
constant and the nonlinearity of the gap equation, a non-spherical
gap function cannot be restricted to one angular momentum channel
only. Other multipoles are bound to show up, which renders the angular
dependence of the gap nontrivial. On the other hand, the pairing strength to the
subleading order decreases with increasing angular momentum $J$.
The mixing effect will not be as big as that in the soluble toy
model we introduced for the purpose of clarification.

For the single flavor CSC, we worked out the angular momentum
mixing effect explicitly for the gap function with zero azimuthal
quantum number at zero temperature. An nonlinear integral equation
for the nontrivial angular dependence was derived and its solution
was obtained numerically. The gap function in this case reads
\begin{equation}
\phi = \left\{
\begin{array}{ll}\Delta_0f(\hat p\cdot\hat z)
\cos\bar{g}{\Big (}\ln\frac{\nu}{\Delta_0|f(\hat p\cdot\hat
z)|}\Big),\hspace{0.3cm}  { \rm for}\hspace{0.2cm}
\nu>\Delta_0|f(\hat p\cdot\hat z)|,\\
\hfill \\
\Delta_0f(\hat p\cdot\hat z),\hspace{3.3cm}  {\rm
for}\hspace{0.2cm} \nu\le\Delta_0|f(\hat p\cdot\hat z)|.
\end{array} \right.
\end{equation}
where $\Delta_0$ is given by Eq.(\ref{delta0})
and $f(\hat p\cdot\hat z)$ is plotted in Fig.\ref{fig:epsart}.

The drop of the free energy of the modified polar phase by the mixing, however, is numerically
small. The magnitude of its condensation energy is smaller than that of the CSL phase
by a factor of 1.48 instead of the factor 1.54 reported in \cite{A}. The CSL phase remains
stable. In this sense our results at the moment is of theoretical
values only.
There are many other candidate pairing states between quarks of the same flavor
\cite{T,A}. Among them are the states with a nonzero azimuthal quantum number
and the pairing between quarks of opposite helicities. The former is analogous to
the $A$ phase of ${\rm He^3}$ and may be present in a compact star with a strong
magnetic field. The pairing force in the unequal-helicity channel is stronger
\cite{T,A,WJH}. The angular momentum mixing effect is generic in all nonspherical
pairing states and the integral equation (\ref{56}) can be readily generalized to
these cases. There may be phenomenological implications of the angular
momentum mixing. A systematic survey of the angular momentum mixing
effect in all "spin-1" CSC states covered in \cite{A} will be
reported in another paper.

Another place where the angular momentum mixing shows up is the
CSC-LOFF state in the presence of Fermi momentum mismatch. It has
been speculated \cite{LRS} that the forward singularity will
increase the upper limit of the mismatch value that supports a
LOFF pairing. The new threshold was found in \cite{GLR}, motivated
by the nearly equal pairing strength of all partial wave channels.
The same mechanism works for the gap equation of LOFF pairing. Its
free energy  will be lowered by the angular momentum mixing and
the lower edge of the LOFF window is expected to be shifted to a
lower value of the mismatch parameter.

%%%%%%%%%%%%%%%%%%%%%

\begin{acknowledgments}

We would like to extend our gratitude to D. Rischke, T.
Sch$\ddot{a}$fer and A. Schmitt
for stimulating discussions and valuable comments. We are also
benefitted from conversations with  J.R. Li and Q. Wang. The work
of D. F. H. and H. C. R. is supported in part by NSFC under grant
No. 10575043 and by US Department of Energy under grants
DE-FG02-91ER40651-TASKB. The work of D. F. H. is also supported in
part by Educational Committee of China under grant NCET-05-0675
and project No. IRT0624

\end{acknowledgments}

\appendix

\section{The derivation of equation (\ref{36})}

The integral formula of $c_J$ is\cite{WJH}
\begin{equation}
c_J=\int\limits_{-1}^{1}dx\frac{P_J(x)-1}{1-x} \label{B1}
\end{equation}
It is convenient to introduce
\begin{equation}
c_J^\epsilon=\int\limits_{-1}^{1}dx\frac{P_J(x)-1}{1-x+\epsilon}
=\int\limits_{-1}^{1}dx\frac{P_J(x)}{1-x+\epsilon}-{\rm
ln}\frac{2+\epsilon}{\epsilon} \label{B2}
\end{equation}
where $\epsilon(>0)$ is an infinitesimal quantity. We have
$\lim_{\epsilon\to 0^+}c_J^\epsilon=c_J$. For the first term on RHS, we expand
\begin{equation}
\frac{1}{1-x+\epsilon}=\sum\limits_{J}a_JP_J(x)
\label{B3}
\end{equation}
according to
\begin{equation}
a_J=\frac{2J+1}{2}\int\limits_{-1}^{1}dx\frac{P_J(x)}{1-x+\epsilon}
\label{B4}
\end{equation}
Therefore Eq. (\ref{B2}) reads
\begin{equation}
c_J^\epsilon=\frac{2}{2J+1}a_J-{\rm ln}\frac{2+\epsilon}{2}
\label{B5}
\end{equation}

Evaluating the summation in Eq.(\ref{36}) is straightforward now
\begin{eqnarray}
\nonumber
\sum\limits_{J=1}^{\infty}(2l+1)c_J^\epsilon P_J({\hat
p}\cdot{\hat
p}^{\prime})&=&\sum\limits_{J=1}^{\infty}2a_JP_J({\hat
p}\cdot{\hat
p}^{\prime})-\sum\limits_{J=1}^{\infty}(2J+1)P_J({\hat
p}\cdot{\hat p}^{\prime}){\rm ln}\frac{2+\epsilon}{\epsilon}\\
&=&\frac{2}{1-{\hat p}\cdot{\hat
p}^{\prime}+\epsilon}-4\pi\delta^2({\hat p}-{\hat p}^{\prime}){\rm
ln}\frac{2+\epsilon}{2}
\label{B6}
\end{eqnarray}
Then for an arbitrary function $f(\hat p)$,
\begin{eqnarray}
\nonumber \int d^2{\hat
p}^{\prime}\sum\limits_{J=1}^{\infty}(2J+1)c_J^\epsilon P_J({\hat
p}\cdot{\hat p}^{\prime})f({\hat p}^{\prime}) &=&2\int d^2{\hat
p}^{\prime}\frac{f({\hat p}^{\prime})}{1-{\hat p}\cdot{\hat
p}^{\prime}+\epsilon}-4\pi f({\hat p}){\rm
ln}\frac{2+\epsilon}{\epsilon}\\
\nonumber
&=&2\int d^2{\hat p}^{\prime}\frac{f({\hat
p}^{\prime})-f({\hat p})}{1-{\hat p}\cdot{\hat
p}^{\prime}+\epsilon}+2\int d^2{\hat p}^{\prime}\frac{f({\hat
p})}{1-{\hat p}\cdot{\hat p}^{\prime}+\epsilon}-4\pi f({\hat
p}){\rm ln}\frac{2+\epsilon}{\epsilon}\\
&=&2\int d^2{\hat p}^{\prime}\frac{f({\hat
p}^{\prime})-f({\hat p})}{1-{\hat p}\cdot{\hat
p}^{\prime}+\epsilon}
\label{B7}
\end{eqnarray}
The Eq. (\ref{36}) is obtained by taking the limit $\epsilon\to 0^+$.

\section{The condensation energy density with the angular momentum mixing}
In this appendix, we shall derive the expression Eq.({\ref{29}})) of the
condensation energy density with the angular momentum mixing.
Substituting Eq.({\ref{33}}) into the first term of Eq. ({\ref{29}}), we find
\begin{eqnarray}
\nonumber F_1&=&-\frac{3{\bar g}^2\mu^4}{32\pi^4}\int
d\nu^{\prime}\int d\nu\int d^2{\hat p^{\prime}}\int d^2{\hat
p}\Big[\frac{1}{6\mu^2}{\rm
ln}\frac{\omega_c}{|\nu-\nu^{\prime}|}\sum_{l=0}^{\infty}(2l+1)P_l(\hat
p\cdot\hat p^{\prime})\\
\nonumber &&+\frac{1}{2\mu^2}\sum_{l=1}^{\infty}(2l+1)c_lP_l(\hat
p\cdot\hat
p^{\prime})\Big]\frac{\phi(\nu,\hat{p})\phi(\nu^{\prime},{\hat
p^{\prime}})}
{\sqrt{(\nu^2+\phi^2(\nu,\hat{p}))(\nu^{\prime2}+\phi^2(\nu^{\prime},{\hat p^{\prime}})})}\\
\nonumber
&=&-\frac{\bar g^2\mu^2}{16\pi^3}\Big\{\int
d\nu^{\prime}\int d\nu\int d^2{\hat p}{\rm
ln}\frac{\omega_c}{|\nu-\nu^{\prime}|}\frac{\phi(\nu,\hat{p})\phi(\nu^{\prime},\hat{p})}
{\sqrt{(\nu^2+\phi^2(\nu,\hat{p}))(\nu^{\prime2}+\phi^2(\nu^{\prime},\hat{p})})}\\
&&-\frac{3}{4\pi}\int d\nu^{\prime}\int d\nu\int d^2{\hat
p^{\prime}}\int d^2{\hat p}\frac{1}{1-{\hat p^{\prime}}\cdot{\hat
p}}\Big[\frac{\phi(\nu, {\hat p})}{\sqrt{\nu^2+\phi^2(\nu, {\hat
p})}}-\frac{\phi(\nu, {\hat p^{\prime}})}{\sqrt{\nu^2+\phi^2(\nu,
{\hat
p^{\prime}})}}\Big]\Big[(\nu\leftrightarrow\nu^{\prime})\Big]\Big\}
\label{A1}
\end{eqnarray}
Because of the eveness of $\phi(\nu,\hat p)$ in $\nu$, we have
\begin{eqnarray}
\nonumber F_0&=&-\frac{\bar g^2\mu^2}{4\pi^3}\Big\{\int d^2{\hat
p}\int\limits_{0}^{\omega_0}d\nu^{\prime}\int\limits_{0}^{\omega_0}{\rm
ln}\frac{\omega_c}{\nu_>}\frac{\phi(\nu, \hat
p)\phi(\nu^{\prime},\hat p)}{\sqrt{[\nu^2+\phi^2(\nu,\hat
p)][\nu^{\prime2}+\phi^2(\nu^{\prime},\hat p)]}}\\
&&-\frac{3}{4\pi}\int d^2{\hat p^{\prime}}\int d^2{\hat
p}\int\limits_{0}^{\omega_0}d\nu^{\prime}\int\limits_{0}^{\omega_0}d\nu\frac{1}{1-{\hat
p}\cdot{\hat p^{\prime}}}\Big[\frac{\phi(\nu, {\hat
p})}{\sqrt{\nu^2+\phi^2(\nu, {\hat p})}}-\frac{\phi(\nu, {\hat
p^{\prime}})}{\sqrt{\nu^2+\phi^2(\nu, {\hat p^{\prime}})}}\Big]
\Big[(\nu\leftrightarrow\nu^{\prime})\Big]\Big\} \label{A3}
\end{eqnarray}
where the approximation (\ref{son}) has been applied to the forward
logarithm. For the gap function of zero azimuthal quantum number,
$\phi(\nu, {\hat p})$ depends only on $x\equiv{\hat p}\cdot{\hat
z}$. We find that
\begin{equation}
F_1=-\frac{\mu^2}{2\pi^2{\bar
g^2}}\Big\{\int\limits_{-1}^{1}dx\int\limits_{a}^{\infty}d\xi\int\limits_{a}^{\infty}d\xi^{\prime}\xi_<
\frac{d\Phi(\xi,x)}{d\xi}\frac{d\Phi(\xi^{\prime},x)}{d\xi^{\prime}}-\frac{3}{2}\int\limits_{-1}^{1}dx\int\limits_{-1}^{1}
\frac{[\Phi(a,x^{\prime})-\Phi(a,x)]^2}{|x-x^{\prime}|}\Big\}
\label{A4}
\end{equation}
where, $\Phi(\xi,x)$ has been defined in Eq.(\ref{38}) and $\xi$ and $a$ have
been defined in (\ref{newvar}). The integral over $\xi^{\prime}$ followed by
the integral by part over $\xi$ leads to
\begin{equation}
F_1=\frac{\mu^2}{2\pi^2{\bar
g^2}}\Big\{-a\int\limits_{-1}^{1}dx\Phi^2(a,x)-\int\limits_{-1}^{1}dx\int\limits_{a}^{\infty}d\xi\Phi^2(\xi,x)+
\frac{3}{2}\int\limits_{-1}^{1}dx\int\limits_{-1}^{1}dx^{\prime}\frac{[\Phi^2(a,x^{\prime})-\Phi^2(a,x)]^2}{|x-x^{\prime}|}\Big\}
\label{A5}
\end{equation}
Making use of Eq. (\ref{38}) and (\ref{53}), we have
\begin{equation}
\int\limits_{a}^{\infty}d\xi\Phi^2(\xi,x)=-{\bar
g^2}\phi_{0}^{2}(x)\Big[\frac{\pi}{2}-{\bar g}(b-a)-{\bar g}{\rm
ln}2\Big]+{\bar
g^2}\int\limits_{a}^{\infty}d\xi\frac{\phi^2(\xi,x)}{\sqrt{1+\frac{\phi^2(\xi,x)}{\omega_c}e^{2\xi}}}
\label{A6}
\end{equation}
 and thus
\begin{equation}
-a\Phi^2(a,x)-\int\limits_{a}^{\infty}d\xi\Phi^2(\xi,x)={\bar
g}\phi_{0}^{2}(x)\Big(\frac{\pi}{2}-{\bar g}b-{\bar g}{\rm
ln}2\Big)-{\bar
g^2}\int\limits_{0}^{\omega_0}d\nu\frac{\phi^2(\nu,x)}{\sqrt{\nu^2+\phi^2(\nu,x)}}
\label{A7}
\end{equation}
Substituting $\phi_0(x)=\Delta_0f(x)$ into (\ref{A5}), we obtain that
\begin{equation}
F_1=\frac{\mu^2\Delta_0^2}{2\pi^2}\Big\{\int\limits_{-1}^{1}dxf^2(x){\rm
ln}|f(x)|+\frac{3}{2}\int\limits_{-1}^{1}dx\int\limits_{-1}^{1}dx^{\prime}\frac{\big[f(x)-f(x^{\prime})\big]^2}
{|x-x^{\prime}|}\Big\}-\frac{\mu^2}{2\pi^2}\int\limits_{0}^{\omega_0}d\nu\frac{\phi^2(\nu,x)}{\sqrt{\nu^2+\phi^2(\nu,x)}}
\label{A8}
\end{equation}
Then the condensate energy density with the angular momentum mixing reads
\begin{eqnarray}
\nonumber F&=&F_1+\frac{\mu^2}{4\pi^3}\int d\nu\int d^2{\hat
p}\Big[|\nu|-\frac{\phi^2(\nu,{\hat
p})}{\sqrt{\nu^2+\phi^2(\nu,{\hat p})}}\Big]\\
&=&\frac{\mu^2\Delta_0^2}{2\pi^2}\Big\{\int\limits_{-1}^{1}dxf^2(x)\Big[{\rm
ln}|f(x)|-\frac{1}{2}\Big]+\frac{3}{2}\int\limits_{-1}^{1}dx\int\limits_{-1}^{1}dx^{\prime}\frac{\big[f(x)-f(x^{\prime})\big]^2}
{|x-x^{\prime}|}\Big\} \label{A9}
\end{eqnarray}
The minimization of this free energy give rise to Eq.(\ref{56}) of the text.

\end{document}